\documentclass[twoside,11pt]{article}

\usepackage{jmlr2e_modified}
\usepackage{amsmath}
\usepackage{makecell}

\usepackage{booktabs}
\usepackage{subcaption}
\usepackage{multirow}

\usepackage[font=small]{caption}

\newcommand{\argmax}{\mathop{\mathrm{argmax}}}

\newcommand{\samplingp}[1]{p_{t,k}^{#1}}

\jmlrheading{ }{ }{}{}{}{}

\ShortHeadings{Methods for Adaptive Experimentation}{Horn and Sloman}
\firstpageno{1}

\begin{document}

\title{A Comparison of Methods for Adaptive Experimentation}

\author{\name Samantha Horn \thanks{Joint first authors.} \email samihorn@cmu.edu \\
       \addr Department of Social and Decision Sciences\\
       Carnegie Mellon University \\
       Pittsburgh, PA 15213 USA
       \AND
       \name Sabina J.\ Sloman \footnotemark[2] \email ssloman@andrew.cmu.edu \\
       \addr Department of Social and Decision Sciences\\
       Carnegie Mellon University\\
       Pittsburgh, PA 15213 USA}

\maketitle

\begin{abstract}%
    We use a simulation study to compare three methods for adaptive experimentation: Thompson sampling, Tempered Thompson sampling, and Exploration sampling. 
    We gauge the performance of each in terms of social welfare and estimation accuracy, and as a function of the number of experimental waves. 
    We further construct a set of novel ``hybrid" loss measures to identify which methods are optimal for researchers pursuing a combination of experimental aims. 
    Our main results are: 1) the relative performance of Thompson sampling depends on the number of experimental waves, 2) Tempered Thompson sampling uniquely distributes losses across multiple experimental aims, and 3) in most cases, Exploration sampling performs similarly to random assignment.
\end{abstract}


\vspace{4mm}

Adaptive experiments have recently gained popularity in the social sciences.\footnote{
    Other literature refers to similar methods as \emph{response-adaptive randomization}.
}
While traditional methods for adaptive experimentation target participant welfare, a body of literature shows that these methods forgo statistical power and can introduce bias in the estimation of the efficacy of some interventions.\footnote{
    See, e.g., \cite{trippa2012bayesian}, \cite{wason2014comparison}, \cite{lin2017comparison}, \cite{wathen2017simulation}, \cite{viele2020comparison}, \cite{ryan2020bayesian} and \citet{ kaibel2021rethinking}.
}
We compare three methods for adaptive experimentation --- Thompson sampling \citep{thompson1933likelihood}, Exploration sampling \citep{kasy2021adaptive} and Tempered Thompson sampling \citep{caria2020adaptive} --- and investigate their relative performance as a function of the number of experimental waves, and with respect to a diverse set of base and hybrid loss measures, corresponding, respectively, to singular and dual experimental aims. 

\section{Problem Setup and Background}
    Consider an experimenter who has access to a population of $N$ experimental participants, each of whom participates in one of $T$ experimental waves, indexed by $t = \{ 1, \ldots{}, T \}$.
    $N_t$ refers to the number of participants who participate in wave $t$.\footnote{
        In our simulations, where $N$ is always evenly divisible by $T$, $N_t = \frac{N}{T} \hspace{1mm} \forall \hspace{1mm} t$.
    }
    We index each participant by $i = \{1, \ldots{}, N_t \}$.
    For each participant $i$ at time $t$, the experimenter observes an outcome $Y_{i,t} \in \{ 0, 1 \}$, with 1 indicating the participant experienced a desirable outcome and 0 indicating the absence of that outcome.
    
    Each participant $i$ at time $t$ is assigned to one of a fixed set of treatments, or interventions, $D_{i,t} \in D$ where $|D| = K$.
    The outcome conditional on reception of treatment $D_k$ is assumed to follow a $Bernoulli(\theta_k)$ distribution.
    $\theta_k$ is the \emph{average potential outcome} corresponding to treatment $D_k$.
    The number of participants assigned to $D_k$ at time $t$ is denoted $n_t^k$. 
    The experimenter starts with a prior distribution on the average potential outcome of each treatment $D_k$.
    After each wave $t$, they use Bayesian inference to update this distribution based on the observed outcomes.
    $p(\theta_k)$ denotes the \emph{posterior} probability of $\theta_k$.
    
    We use $k^*$ to index the treatment with the highest average potential outcome (unknown to the experimenter).
    $\hat{k}$ indexes the treatment with the highest estimated average potential outcome at the end of the experiment, i.e., $\hat{k} \equiv \argmax_{k \in \{ 1, \ldots{}, K \}} \int_{\theta_k} \theta_k \hspace{1mm} p(\theta_k) \hspace{1mm} d\theta_k$.
    In practice, this can be thought of as the treatment deemed most likely to be effective based on the data collected, and perhaps implemented as policy.

\section{Description of Assignment Mechanisms}\label{sec:assignment-mechanisms}
    Each adaptive experimentation method, or assignment mechanism, we evaluate differs in how $n_t^k$, the number of participants assigned to each treatment $D_k$ at wave $t$, is determined. 
    We compare all assignment mechanisms to the baseline of \emph{random assignment (RA)} in which the probability of assignment to each treatment is constant across waves and is simply $\frac{1}{K}.$

    When using \emph{Thompson sampling} \citep{thompson1933likelihood}, the probability of assignment to treatment group $k$ in experimental wave $t$ is:
        
        $$\samplingp{thompson} = \mathbb{P}(k = k^*)$$ 
        
    \emph{Exploration sampling} \citep{kasy2021adaptive} provides a slight modification to Thompson sampling and is designed to increase power for rejecting suboptimal treatments. 
    This is achieved by modifying the assignment probabilities as follows:

        $$\samplingp{exploration} = \frac{\samplingp{thompson}(1-\samplingp{thompson})}{\sum_k \samplingp{thompson}(1-\samplingp{thompson})}$$

    \emph{Tempered Thompson sampling} is a method intended to strike a balance between painting an overall picture of the effectiveness of each treatment and minimizing in-sample regret \citep{caria2020adaptive}.
    It assigns participants to arm $k$ proportionally to the weighted average of $\frac{1}{K}$ (the assignment probability under RA) and $p_{t,k}^{thompson}$.
    In other words, the probability of assignment to treatment group $k$ in experimental wave $t$ is:
        
        $$\samplingp{tempered}  = (1-\gamma)\samplingp{thompson} + \frac{\gamma}{K}$$
        
    where $\gamma\in[0,1]$ allows researchers a degree of freedom in how much weight is placed on the Thompson assignment probabilities.
    $\gamma$ can also be thought of as controlling how much the sampling process targets regret minimization over estimation accuracy.\footnote{
        In our simulations, we set $\gamma = .2$.
    }
    
\section{Experimental Setup}
    Each of our simulated experiments tested three ``treatments," each with a true average potential outcome drawn from a standard uniform distribution.
    For each set of three treatments, we ran experiments using each of the four assignment mechanisms described above at each of three levels of $N_t$: $N_t \in \{ 4, 10, 100 \}$.
    For each experiment we fixed the total population size $N$ at $1,000$, in effect predetermining the number of experimental waves, $T \in \{ 250, 100, 10 \}$.
    We thus ran 4 assignment mechanisms $\times$ 3 levels of $N_t$ $\times$ 10,000 sets of treatments = 120,000 experiments in total.
    At the beginning of each experiment, we began with an uninformative $Beta(1,1)$ prior for each of $\theta_1$, $\theta_2$ and $\theta_3$.\footnote{
        Replication code available at https://github.com/sami-horn/adaptive-experimentation.
    }
    
    \paragraph{Loss measures.}
        For each experiment, we analyze its performance with respect to several loss measures, each of which corresponds to a potential experimental goal.
        Table \ref{tab:loss} summarizes the three classes of loss measures we consider: measures of \emph{regret}, \emph{estimation precision} and \emph{statistical power}.
        
        \begin{table}[]
            \renewcommand{\arraystretch}{1.5}
            \centering
            \begin{tabular}{c|c|c|c}
                \midrule
                \multicolumn{1}{c}{} & \multicolumn{1}{c}{\bf Description} & \multicolumn{1}{c}{\bf Notation} & \multicolumn{1}{c}{\bf Calculation} \\ \midrule
                \multirow{2}{*}{Regret} & In-sample regret & $R_{sample}$ & $\frac{1}{N} \sum_{i = 1}^T \sum_{i = 1}^{N_t} \Delta_{D_{i,t}}$ \\
                 & Policy regret & $R_{policy}$ & $\Delta_{D_{\hat{k}}}$ \\ \midrule
                \multirow{1.5}{*}{Estimation} & RMSE of $\theta_{\hat{k}}$ & $PREC_{best}$ & $RMSE_{\hat{k}}$ \\
                \multirow{.8}{*}{precision} & Average RMSE & $PREC_{avg}$ & $\frac{1}{K}\sum_{k=1}^K RMSE_k$ \\[1ex] \midrule
                \multirow{1.5}{*}{Statistical} & Fails to order  & \multirow{2.5}{*}{$SP$} & \multirow{2.5}{*}{$1- \mathbb{I}\left(\mathbb{P}\left(\hat\theta_{(k)} >  \hat\theta_{(k-1)} \right) < \alpha  \; \forall k \in \{ 2,\ldots, K \} \right) $}\\[-1.5ex]
                \multirow{1.5}{*}{power} & treatments by & &  \\[-1.5ex]
                 & $\theta_k$& & \\[1ex]
                 \midrule
            \end{tabular}
            \caption{Loss measures.}
            \label{tab:loss}
        \vspace{-3mm}
        \end{table}

        \emph{Regret-based measures} rely on the \emph{regret} $\Delta_{D_k}$ associated with a particular treatment $D_k$.
        Regret measures the amount of welfare lost compared to what would have been lost if all receivers were assigned to $D_{k^*}$.
        Formally, it is defined as
        \begin{equation*}
            \Delta_{D_k} \equiv \theta^*_{k^*} - \theta^*_k
        \end{equation*}
        where $\theta^*_k$ indicates the true (in practice, unknowable) effect of treatment $D_k$.
        
        \emph{Precision-based measures} rely on the \emph{root mean-squared error} ($RMSE_k$) of the posterior distribution of the average potential outcome associated with a particular $D_k$:
        \begin{equation*}
            RMSE_k \equiv \sqrt{\int_{\theta_k} (\theta^*_k - \theta_k)^2 \hspace{1mm} p(\theta_k) \hspace{1mm} d \theta_k}
        \end{equation*}
        
        Our \emph{power-based measure} determines whether the study was able to identify the correct ordering of arms based on their true average potential outcomes.
        It measures the ability of a series of statistical tests with controlled Type-I error to recover the true rank order of $\theta^*_1$, $\theta^*_2$ and $\theta^*_3$.\footnote{
            In our empirical results, we fix the Type-I error for each pairwise hypothesis test to .05, and use Monte Carlo draws from each $p(\theta_k)$ to generate empirical $p$-values.
        }
        
        \emph{Hybrid loss measures} are pairwise combinations of the ``base" loss measures described above.
        For example, a hybrid of $R_{sample}$ and $PREC_{avg}$ (denoted by $R_{sample}/PREC_{avg}$) would represent the dual goal of both maximizing social welfare in the participant sample and the precision of the estimated average potential outcomes.
        Because the regret- and precision-based measures are computed on the same scale (each corresponds to the magnitude of a difference between two average potential outcomes\footnote{
            In the case of the precision measures, this is the expectation of a difference with respect to the posterior distribution of the average potential outcome.
        } and is lower-bounded by 0 and upper-bounded by 1), for hybrid loss measures that are made up of combinations of a regret and precision loss measure we simply take the average of the two measures.
        For hybrid loss measures that combine a regret- or precision-based measure $L$ with $SP$, we take the maximum value of the two measures.
        This equals the value of $L$ in case the correct ordering is identified ($SP = 0$); otherwise, the maximum loss of 1 is incurred. 
        This can be interpreted similarly to a constrained objective, in which the ``constraint" is that the correct ordering is identified.

\section{Results}
\begin{figure}[h]
            \centering
            \includegraphics[width=\textwidth]{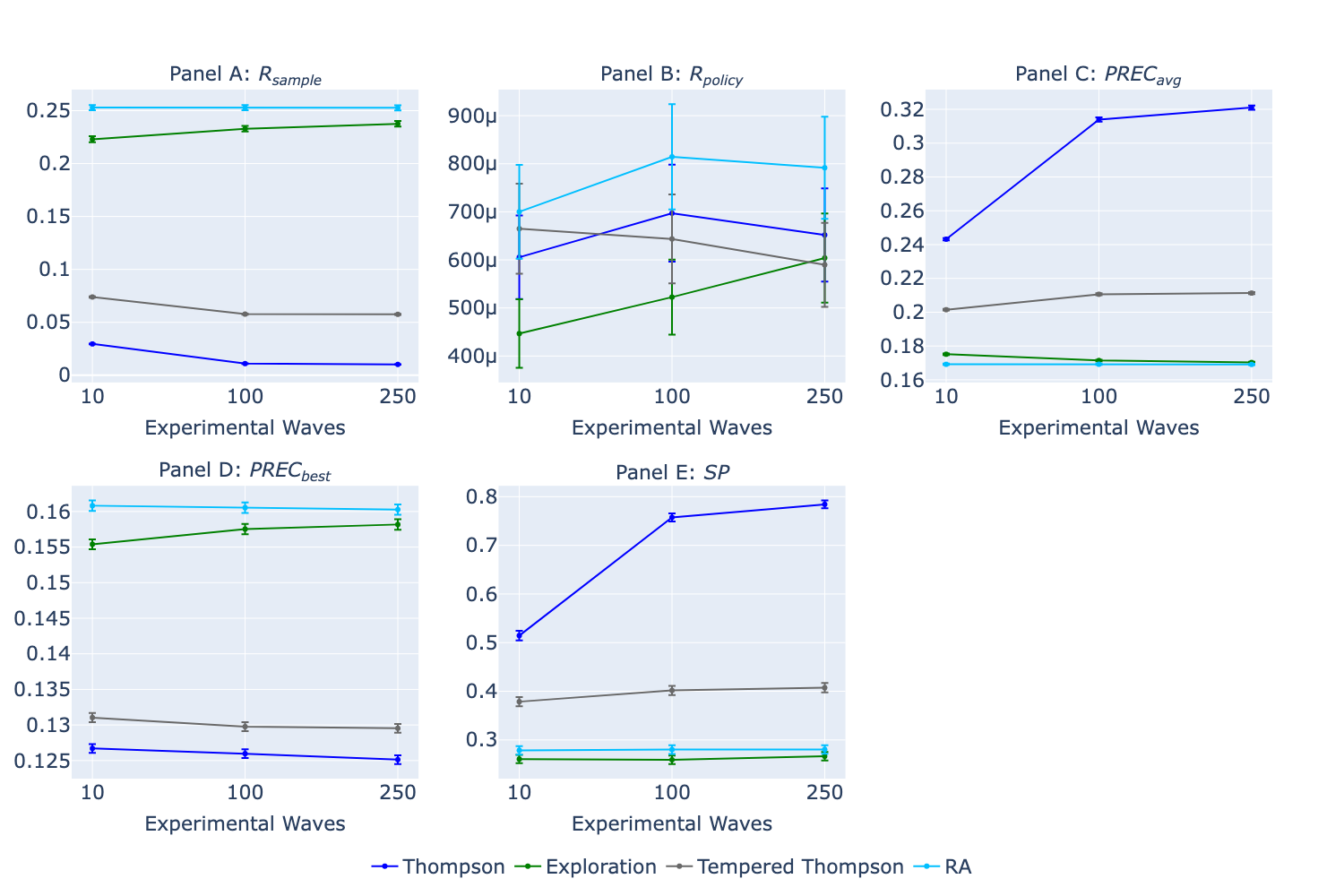}
            \caption{Average performance on loss measures as a function of number of experimental waves. See Section 4.1 for details on loss measures. Error bars represent 95\% confidence intervals.}
            \label{fig:fig1}
        \end{figure}

    \paragraph{Base loss measures.}
        Panels A and B of Figure \ref{fig:fig1} show performance on the two \textit{regret-based measures}. 
        $R_{sample}$ is minimized by Thompson sampling regardless of the number of experimental waves. 
        $R_{policy}$ is generally imprecisely measured and very low, suggesting that all methods usually identify the best treatment arm.

        Panels C and D show how each method performs on the two \textit{precision-based measures}. 
        Thompson sampling results in higher $PREC_{avg}$ than other methods, and the $PREC_{avg}$ values associated with Thompson sampling increase dramatically with the number of experimental waves.
        The pattern of results for $PREC_{best}$ is similar to $R_{sample}$.

        Finally, Panel E plots performance for $SP$. 
        This resembles the patterns shown in Panel C, which reflects that both $PREC_{avg}$ and $SP$ require precise estimation of the average potential outcomes associated with all three treatments.
        However, Exploration sampling consistently outperforms RA on $SP$.

        Overall, Tempered Thompson sampling performs similarly to or better than Thompson sampling, without exhibiting large variation in performance by the number of experimental waves.
    
\begin{figure}[]
            \includegraphics[width=\linewidth]{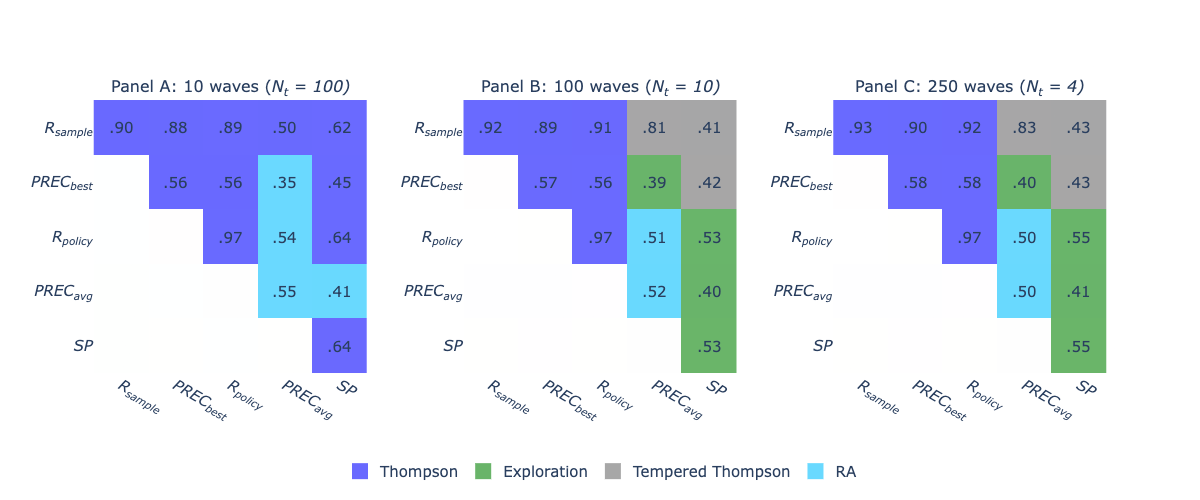}
	        \caption{
	            The assignment mechanism that most often minimizes each of the hybrid loss measures (the diagonal indicates the loss-minimizing assignment mechanism for each base loss measure).
	            Numbers indicate the proportion of simulations on which the indicated assignment mechanism had the lowest corresponding loss.
	        }
            \label{fig:hybrid_overall}
        \end{figure}

    \paragraph{Hybrid loss measures.}
        Figure \ref{fig:hybrid_overall} shows the loss-minimizing assignment mechanism for each possible hybrid measure.
        To identify the ``loss-minimizing" mechanism, we computed the hybrid loss achieved by each assignment mechanism on each experiment, and identified the mechanism which achieved the lowest loss on the greatest number of trials.\footnote{
            We ran a similar analysis treating the loss-minimizer as the mechanism achieving the lowest average loss across experiments.
            Those results differ from those shown here in two notable ways: 1) Panel A resembles Panels B and C, i.e., Thompson sampling's advantages when there are few experimental waves are not apparent, and 2) Thompson sampling is never selected as the loss-minimizer for $R_{policy}$ (as shown in Panel B of Figure \ref{fig:fig1}, on this measure, Thompson sampling is outperformed at all levels of $N_t$).
        }

        When the number of experimental waves is small, Thompson sampling most often minimizes loss according to almost every measure, outperformed by RA on only $PREC_{avg}$, $PREC_{avg} / PREC_{best}$, $PREC_{avg} / R_{policy}$ and $PREC_{avg} / SP$ --- all of which require accurate estimation of the average potential outcomes of all treatment arms.

        However, this seemingly near-universal benefit of Thompson sampling does not persist in the case of large numbers of experimental waves.
        In these cases, Thompson sampling performs well for pairwise combinations of $R_{sample}$, $PREC_{best}$ and $R_{policy}$.
        In a complementary pattern, Exploration sampling and RA perform well for pairwise combinations of $R_{policy}$, $PREC_{avg}$ and $SP$.
        Further inspection showed that, with the exception of $SP/R_{policy}$ and $SP$\footnote{
            We discuss Exploration sampling's persistent advantage with respect to $SP$ above; since the values of $R_{policy}$ are so small, $SP/R_{policy}$ is usually dominated by $SP$.
        }, Exploration sampling and RA perform similarly on all of these measures, highlighting the ability of both to accurately estimate the average potential outcomes of all treatments.

        Our results suggest that Tempered Thompson sampling is best when the objective requires \emph{both} over-sampling from the best treatment ($R_{sample}$ and $PREC_{best}$) \emph{and} precise estimates for all treatment arms ($SP$ and $PREC_{avg}$).
        Notably, Tempered Thompson sampling does not excel at minimizing any base measure in isolation; its comparative advantage stems from its ability to distribute losses across dual experimental aims.
        This reflects the fact that Thompson sampling is constructed as a blend of two other assignment mechanisms, Thompson sampling and RA, with the explicit aim of striking a balance between the benefits of both (see section \ref{sec:assignment-mechanisms}).
    
\section{Discussion}
    We evaluated three methods for adaptive experimentation with respect to a set of base and hybrid loss measures.
    We found that 1) the relative performance of Thompson sampling depends on how participants are distributed across experimental waves, 2) Exploration sampling maximizes statistical power to discriminate between treatment arms \citep{kasy2021adaptive}, and 3) Tempered Thompson sampling balances overall statistical power with an understanding of the apparently best treatment \citep{caria2020adaptive}.
    
    While our hybrid loss measures represent one way of constructing a quantitative trade-off between dual experimental aims, more practically useful measures would attribute weight to different aims in a way that more closely reflects the objectives of a particular researcher or problem domain.
    Construction of such application-specific measures is an important next step for future work.
    
\newpage

\acks{We would like to acknowledge support for this work from the Center for Machine Learning and Health (CMLH) at Carnegie Mellon University. 
SJS was supported by a Tata Consultancy Services (TCS) Fellowship while contributing to this work.
}

\vskip 0.2in
\bibliography{adaptive_experimentation}

\begin{thebibliography}{10}
\providecommand{\natexlab}[1]{#1}
\providecommand{\url}[1]{\texttt{#1}}
\expandafter\ifx\csname urlstyle\endcsname\relax
  \providecommand{\doi}[1]{doi: #1}\else
  \providecommand{\doi}{doi: \begingroup \urlstyle{rm}\Url}\fi

\bibitem[Caria et~al.(2020)Caria, Kasy, Quinn, Shami, Teytelboym,
  et~al.]{caria2020adaptive}
Stefano Caria, Maximilian Kasy, Simon Quinn, Soha Shami, Alex Teytelboym,
  et~al.
\newblock An adaptive targeted field experiment: Job search assistance for
  refugees in jordan.
\newblock 2020.

\bibitem[Kaibel and Biemann(2021)]{kaibel2021rethinking}
Chris Kaibel and Torsten Biemann.
\newblock Rethinking the gold standard with multi-armed bandits: Machine
  learning allocation algorithms for experiments.
\newblock \emph{Organizational Research Methods}, 24\penalty0 (1):\penalty0
  78--103, 2021.

\bibitem[Kasy and Sautmann(2021)]{kasy2021adaptive}
Maximilian Kasy and Anja Sautmann.
\newblock Adaptive treatment assignment in experiments for policy choice.
\newblock \emph{Econometrica}, 89\penalty0 (1):\penalty0 113--132, 2021.

\bibitem[Lin and Bunn(2017)]{lin2017comparison}
Jianchang Lin and Veronica Bunn.
\newblock Comparison of multi-arm multi-stage design and adaptive randomization
  in platform clinical trials.
\newblock \emph{Contemporary clinical trials}, 54:\penalty0 48--59, 2017.

\bibitem[Ryan et~al.(2020)Ryan, Lamb, Williamson, and Gates]{ryan2020bayesian}
Elizabeth~G Ryan, Sarah~E Lamb, Esther Williamson, and Simon Gates.
\newblock Bayesian adaptive designs for multi-arm trials: an orthopaedic case
  study.
\newblock \emph{Trials}, 21\penalty0 (1):\penalty0 1--16, 2020.

\bibitem[Thompson(1933)]{thompson1933likelihood}
William~R Thompson.
\newblock On the likelihood that one unknown probability exceeds another in
  view of the evidence of two samples.
\newblock \emph{Biometrika}, 25\penalty0 (3-4):\penalty0 285--294, 1933.

\bibitem[Trippa et~al.(2012)Trippa, Lee, Wen, Batchelor, Cloughesy, Parmigiani,
  and Alexander]{trippa2012bayesian}
Lorenzo Trippa, Eudocia~Q Lee, Patrick~Y Wen, Tracy~T Batchelor, Timothy
  Cloughesy, Giovanni Parmigiani, and Brian~M Alexander.
\newblock Bayesian adaptive randomized trial design for patients with recurrent
  glioblastoma.
\newblock \emph{Journal of Clinical Oncology}, 30\penalty0 (26):\penalty0 3258,
  2012.

\bibitem[Viele et~al.(2020)Viele, Broglio, McGlothlin, and
  Saville]{viele2020comparison}
Kert Viele, Kristine Broglio, Anna McGlothlin, and Benjamin~R Saville.
\newblock Comparison of methods for control allocation in multiple arm studies
  using response adaptive randomization.
\newblock \emph{Clinical Trials}, 17\penalty0 (1):\penalty0 52--60, 2020.

\bibitem[Wason and Trippa(2014)]{wason2014comparison}
James~MS Wason and Lorenzo Trippa.
\newblock A comparison of bayesian adaptive randomization and multi-stage
  designs for multi-arm clinical trials.
\newblock \emph{Statistics in medicine}, 33\penalty0 (13):\penalty0 2206--2221,
  2014.

\bibitem[Wathen and Thall(2017)]{wathen2017simulation}
J~Kyle Wathen and Peter~F Thall.
\newblock A simulation study of outcome adaptive randomization in multi-arm
  clinical trials.
\newblock \emph{Clinical Trials}, 14\penalty0 (5):\penalty0 432--440, 2017.

\end{thebibliography}

\end{document}